\documentclass[12pt]{article}

\usepackage{amsfonts,amssymb,amsmath}

\usepackage{graphicx}
\DeclareGraphicsExtensions{.pdf,.eps,.jpg}

\begin{document}
\title{\bf Quantum speedup, non-Markovianity and formation of bound state}
\author{ Bahram Ahansaz
\thanks{E-mail:bahramahansaz@gmail.com}
and Abbas Ektesabi
\thanks{E-mail:ektesabi.abbas@gmail.com}
\\ {\small Physics Department, Azarbaijan Shahid Madani University, Tabriz, Iran}}\maketitle

\begin{abstract}
\noindent
In this paper, we investigate the relationship between the quantum speedup, non-Markovianity and formation of a system-environment bound state.
Previous results show a monotonic relation between these three such that providing stronger bound states can lead to a higher degree of non-Markovianity, and hence to greater speed of quantum evolution. By studying dynamics of a dissipative two-level system or a V-type three-level system in the presence of similar and additional systems, we reveal that the quantum speedup is exclusively related to the formation of the system-environment bound state, while the non-Markovian effect of the system dynamics is neither necessary nor sufficient to speed up the quantum evolution. In the other hand, it is demonstrated that only the formation of the system-environment bound state plays a decisive role in the acceleration of the quantum evolution.
\end{abstract}

\section{Introduction}
The quantum speed limit (QSL) time $\cite{Mandelstam, Margolus}$ is defined as the minimal time a quantum system needs to evolve from an initial state to a target state, which can be used to characterize the maximal speed of evolution of a quantum system. It arises in tremendous areas of quantum physics and quantum information, such as nonequilibrium thermodynamics $\cite{Lutz}$, quantum metrology $\cite{Giovannetti}$, quantum optimal control $\cite{Caneva}$, quantum computation $\cite{Bekenstein,Lloyd1}$, quantum communication $\cite{Lutz,Yung}$,
and has been studied extensively due to its fundamental importance. The problem of determining  how to exploit the available resources to achieve the highest evolution speed is relevant to deriving physical limits in a variety of contexts. In closed systems, entanglement has been taken as a resource in the speedup of quantum evolution $\cite{Batle,Borras,Frowis}$.
Since any real system is coupled to its environment in the practical scenario, a considerable amount of work has witnessed research on controlling speedup in open quantum systems. In this regard, Deffner and Lutz showed that the non-Markovian process induced by the memory effect of environment can induce dynamical acceleration in the damped Jaynes-Cummings model for a two-level system, and therefore lead to a smaller QSL time $\cite{Deffner}$. This phenomenon has been verified in different settings $\cite{Mo,Zhang,Xu}$ and confirmed by the experiment in cavity quantum electrodynamics (QED) systems $\cite{Cimmarusti}$, where an atomic beam is treated as a controllable environment for the cavity field system. Recently, the authors of Ref. $\cite{Liu}$ showed that both the non-Markovianity and the quantum speedup are attributed to the formation of a system-environment bound state, i.e., an eigenstate of the system-reservoir Hamiltonian with negative eigenenergy which is separated from the other eigenstates by an energy gap $\cite{Yablonovitch,John,Zhu,Tong}$. Via considering a two-level atom coupled to an environment with Ohmic spectrum, it was demonstrated that if the bound state is established, so the evolution of the system becomes non-Markovian, and thus quantum speedup occurs. They found a monotonic relation between them such that providing stronger bound states can lead to a higher degree of non-Markovianity, and hence to greater speed of quantum evolution. Despite the growing body of literature on this subject, the analysis has almost exclusively focused on the simpler settings and to our knowledge, the mechanism of speedup has not been investigated in the more complex settings.

The purpose of this paper is to investigate the relationship between the quantum speedup, non-Markovianity and formation of a system-environment bound state in the more complex settings. To this aim, we study the mechanism of quantum speedup for a two-level system and a V-type three level system in the presence of similar and additional systems. We illustrated that there is a monotonic relation between the quantum speedup and the formation of a system-environment bound state. In particular, we show that the stronger bound states provided by the additional systems can speed up the quantum evolution. However, it is figure out that there is not such monotonic relation between the quantum speedup and the non-Markovianity. We demonstrate that although non-Markovianity governs the quantum speedup in the absence of additional systems,
but it is neither necessary nor sufficient to speed up the quantum evolution when the environment become more complex in the presence of additional systems. In the other hand, our results suggest that only the formation of the system-environment bound state is the essential reason for the quantum speedup.

\section{The condition for the existence of a bound state in a system including $N$ atoms}
In this section, we consider a system containing $N$ non-interacting atoms in a common reservoir, as depicted in Fig. 1.
As will be discussed in the following subsections, we check the energy spectrum of the Hamiltonian of the mentioned system by considering the atoms as two-level and three-level systems.
Then, it will be shown that how the system-environment bound state, as special eigenstate with eigenvalue residing in the band gap of the energy spectrum, can be formed in these models.

\subsection{Two-level systems}
By considering the atoms as two-level systems, the Hamiltonian of the whole system can be written as ($\hbar=1$)
\begin{eqnarray}
H=\omega_{0} \sum_{l=1}^{N} \sigma^{+}_{l} \sigma^{-}_{l}+\sum_{k} \omega_{k} b_{k}^{\dagger} b_{k}+\sum_{l=1}^{N} \sum_{k} (g_{k} b_{k} \sigma_{l}^{+}+g_{k}^{*} b_{k}^{\dagger} \sigma_{l}^{-}),
\end{eqnarray}
where the first term is the Hamiltonian of $N$ two-level systems, the second term describes the reservoir Hamiltonian and the last terms is the system-reservoir interaction Hamiltonian.
Also, $\sigma_{l}^{+}$ $(\sigma_{l}^{-})$ is the raising (lowering) operator of the $l$th atom with transition frequency $\omega_{0}$ and $b_{k}^{\dagger}$ ($b_{k}$) is the creation (annihilation) operator of the $k$th field mode with frequency $\omega_{k}$ and $g_{k}$ is the strength of coupling between the $l$th atom and the $k$th field mode.

The energy spectrum of the total Hamiltonian, $H$, can be obtained by solving the following eigenvalue equation
\begin{eqnarray}
 H |\psi(t)\rangle=E |\psi(t)\rangle.
\end{eqnarray}
Since the total Hamiltonian commutes with the number of excitations (i. e. $[(\sum_{l=1}^{N} {\sigma_{l}^{+} \sigma_{l}^{-}}+\sum_{k} b_{k}^{\dagger} b_{k}),H]=0$), therefore by considering the single excitation subspace, the state of the whole system can be assumed to be as follows
\begin{eqnarray}
  |\psi(t)\rangle=\alpha_{0}(t)|0\rangle_{S} |0\rangle_{E}+\sum_{l=1}^{N}\alpha_{l}(t)|l\rangle_{S} |0\rangle_{E} +\sum_{k}\beta_{k}(t)|0\rangle_{S} |1_{k}\rangle_{E},
\end{eqnarray}
where $|l\rangle_{S}=|g\rangle^{\bigotimes N}_{l\mathrm{th}\equiv e}$, which means that all of the atoms are in their respective ground states $|g\rangle$ except the $l$th atom which is in the excited state $|e\rangle$, and  $|0\rangle_{S}=|g\rangle^{\bigotimes N}=|g,g,...,g\rangle$.
Also, we denote $|0\rangle_{E}$ being the vacuum state of the reservoir and $|1_{k}\rangle_{E}$ is the state for which there is only one excitation in the $k$th field mode.
Now, we substitute Eqs. (1) and (3) into Eq. (2) and obtain the following set of $N+1$ equations
\begin{eqnarray}
\begin{array}{c}
  \omega_{k} \beta_{k}(t)+\sum_{l=1}^{N} g_{k}^{*} \alpha_{l}(t)=E \beta_{k}(t),\\\\
  \omega_{0} \alpha_{j}(t)+\sum_{k} g_{k} \beta_{k}(t)=E \alpha_{j}(t),  \quad j=1, 2,...,N.
\end{array}
\end{eqnarray}
Obtaining $\beta_{k}(t)$ from the first equation and substituting it in the rest ones leads to the following $N$ integro-differential equations
\begin{eqnarray}
  (E-\omega_{0}) \alpha_{j}(t)=-\int_{0}^{\infty} \frac{J(\omega) d\omega}{\omega-E} \sum_{l=1}^{N} \alpha_{l}(t),  \quad j=1, 2,...,N.
\end{eqnarray}
After summing the above $N$ equations and eliminating $\sum_{l=1}^{N} \alpha_{l}(t)$ from both sides of it, we obtain
\begin{eqnarray}
\mathcal{K}(E)=E,
\end{eqnarray}
with
\begin{eqnarray}
\mathcal{K}(E)=\omega_{0}-N \int_{0}^{\infty} \frac{J(\omega) d\omega}{\omega-E},
\end{eqnarray}
which depends on the particular choice of the spectral density $J(\omega)$ of the reservoir and the number of atoms $N$ in the reservoir.
A bound state is an eigenstate with real (discrete) eigenvalue in a quantum many-body system. So, we can claim that the system possesses a bound state, when Eq. (6) has a real root.
It can be easily seen that $\mathcal{K}(E)$ decreases monotonically with the increase of $E$ in the regime of $E <0$. Therefore, $\mathcal{K}(E)$ always has one and only one intersection with the function on the right-hand side of Eq. (6) when the condition $\mathcal{K}(0) < 0$ is satisfied.
This root just corresponds to the eigenvalue of the formed bound state in the Hilbert space of the whole system ($N$ two-level system plus its common reservoir).
Moreover, in the regime of $E >0$, we can observe that $\mathcal{K}(E)$ is divergent and no real root can make Eq. (7) well defined.
Consequently, Eq. (6) does not have a real root to support the existence of a further bound state in this regime.

\subsection{Three-level systems}
After studying the energy spectrum of the system including $N$ two-level systems in a common reservoir, we investigate this problem by considering three-level systems in V configuration.
In the other hand, we assume that each atom has two excited states and each of which spontaneously decays into ground state such that the respective dipole moments of transitions may have interaction with each other through spontaneously generated interference (SGI). Also, for a given V-type atom, an excited states are characterized by $|A\rangle$ and $|B\rangle$ which can spontaneously decay into ground state $|C\rangle$ with transition frequencies $\omega_{A}$ and $\omega_{B}$, respectively. The Hamiltonian for this described system can be written as
\begin{eqnarray}
H'=\sum_{l=1}^{N} \sum_{m=A,B} \omega_{m} \sigma_{m}^{l+} \sigma_{m}^{l-} +\sum_{k}\omega_{k}b_{k}^{\dagger}b_{k}+\sum_{l=1}^{N} \sum_{m=A,B} \sum_{k} \big(g_{mk} \sigma_{m}^{l+} b_{k}+g_{mk}^{\ast} \sigma_{m}^{l-} b_{k}^{\dagger} \big).
\end{eqnarray}
Here, ${\sigma_{m}^{l\pm}}(m=A, B)$ are the raising and lowering operators of the $m$th excited state to the ground state for the $l$th atom.
Also, the annihilation (creation) operator of the $k$th field mode is given by $b_{k}$ ($b_{k}^{\dagger}$) with the frequency $\omega_{k}$
and the strength of coupling between the $m$th excited state and the $k$th field mode for all atoms is assumed to be identical and is given by $g_{mk}$.

Now, we want to study the energy spectrum of $N$ three-level systems in a common reservoir. So, we solve the following eigenvalue equation
\begin{eqnarray}
H'|\varphi(t)\rangle=E'|\varphi(t)\rangle,
\end{eqnarray}
where
\begin{eqnarray}
  |\varphi(t)\rangle=\nu_{0}(t)|0\rangle_{S} \otimes |0\rangle_{E}+\sum_{l=1}^{N} \big( \nu_{l}^{A}(t)|A_{l}\rangle+ \nu_{l}^{B}(t)|B_{l}\rangle \big)_{S} \otimes |0\rangle_{E}
  +\sum_{k} \eta_{k}(t) |0\rangle_{S} |1_{k}\rangle_{E},
\end{eqnarray}
is the state of the whole system in the single excitation subspace, because the total Hamiltonian conserves the number of excitations in the system,
i.e. $[\sum_{l=1}^{N}(\sigma_{A}^{l+}\sigma_{A}^{l-}+\sigma_{B}^{l+} \sigma_{B}^{l-})+\sum_{k} b_{k}^{\dagger} b_{k},H']=0$.
Also, $|0\rangle_{S}=|C\rangle^{\otimes N}$ means that all of the atoms are in the ground state and $|A_{l}\rangle_{S}$ and $|B_{l}\rangle_{S}$ represent that all of the atoms are in the ground state except the $l$th atom which is respectively in the first and second upper excited levels. We denote $|0\rangle_{E}$ is the vacuum state of the reservoir and $|1_{k}\rangle_{E}$ the state of it with only one excitation in the $k$th field mode. For simplicity, we assume the case of the degenerate excited levels where $\omega_{A}=\omega_{B}=\omega_{0}$.
Then, by substituting Eqs. (8) and (10) into Eq. (9), we obtain the following differential equations
$$
\omega_{k} \eta_{k}(t)+\sum_{l=1}^{N} \big(g_{Ak}^{*} \nu_{l}^{A}(t)+g_{Bk}^{*} \nu_{l}^{B}(t)\big) =E \eta_{k}(t),
$$
\begin {equation}
\omega_{0} \nu_{l}^{A}(t)+\sum_{k} g_{Ak} \eta_{k}(t)=E \nu_{l}^{A}(t), \quad l=1, 2,...,N,
\end {equation}
$$
\omega_{0} \nu_{l}^{B}(t)+\sum_{k} g_{Bk} =E \nu_{l}^{B}(t), \quad l=1, 2,...,N.
$$
We derive $\eta_{k}(t)$ from the first equation and substitute it into the rest ones, so we have
\begin{equation}
\begin{array}{l}
\displaystyle (E-\omega_{0}) \nu_{l}^{A}(t)=-\int_{0}^{\infty} \frac{J(\omega) d\omega}{\omega-E} \sum_{l=1}^{N} \nu_{l}^{A}(t)-\int_{0}^{\infty} \frac{J'(\omega) d\omega}{\omega-E} \sum_{l=1}^{N} \nu_{l}^{B}(t),\\\\
\displaystyle (E-\omega_{0}) \nu_{l}^{B}(t)=-\int_{0}^{\infty} \frac{J'(\omega) d\omega}{\omega-E} \sum_{l=1}^{N} \nu_{l}^{A}(t)-\int_{0}^{\infty} \frac{J(\omega) d\omega}{\omega-E} \sum_{l=1}^{N} \nu_{l}^{B}(t).
\end{array}
\end{equation}
Finally, a compact form of these equations can be obtained by eliminating the amplitudes coefficients as
\begin{eqnarray}
\mathcal{K}'(E')=E',
\end{eqnarray}
where
\begin{eqnarray}
\mathcal{K}'(E')=\omega_{0}-N(1+\theta) \int_{0}^{\infty} \frac{J(\omega) d\omega}{\omega-E'}.
\end{eqnarray}
It is clear that the solutions of Eq. (13) depend on the particular choice of the spectral density of the reservoir $J(\omega)$, the number of atoms $N$ and the SGI parameter $\theta$.
Nevertheless, as previously discussed in the last subsection, the bound state is formed when Eq. (13) have at least a real solution in the negative energy range, i.e., $E <0$, and otherwise the bound state is not formed.

\section{QSL time and non-Markovianity}
In this section, we want to evaluate QSL time bound and non-Markovianity of a single atom (two-level or three-level) in the presence of $N-1$ additional atoms.
In this regard, the first atom ($l$=1) of the studied system in section II, is considered as our main concern of the single system and the $N-1$ remainder ones are
considered as the additional atoms.

A unified expression for the QSL time in open systems, widely used to evaluate the speed of quantum evolution, in given by $\cite{Deffner}$
\begin{eqnarray}
\tau_{\mathrm{QSL}}=\mathrm{max} \{\frac{1}{\Lambda^{\mathrm{1}}_{\tau}},\frac{1}{\Lambda^{\mathrm{2}}_{\tau}},\frac{1}{\Lambda^{\mathrm{\infty}}_{\tau}}\} \mathrm{sin}^2\big[\mathcal{B}(\rho(0),\rho(\tau))\big],
\end{eqnarray}
where $\mathcal{B}(\rho(0),\rho(\tau))=\mathrm{arccos}\Big(\sqrt{\langle\phi(0)|\rho(\tau)|\phi(0)\rangle}\Big)$, is the Bures angle between initial pure state
$\rho(0)=|\phi(0)\rangle \langle \phi(0)|$ and the target state $\rho(\tau)$ governed by the time-dependent non-unitary equation $\mathcal{L}_{t}(\rho(t))=\dot{\rho}(t)$, and
\begin{eqnarray}
\Lambda^{\mathrm{p}}_{\tau}=\frac{1}{\tau}\int_{0}^{\tau}dt\|\mathcal{L}_{t}(\rho(t))\|_{\mathrm{p}},
\end{eqnarray}
with $\|X\|_{\mathrm{p}}=(x_{1}^{p}+...+x_{n}^{p})^{1/p}$, which denotes the Schatten p-norm and $x_{1},...,x_{n}$ are the singular values of $X$. Using this QSL time bound enables us to evaluate the intrinsic speed of the dynamical evolution by a given actual driving time $\tau$. On the other hand, when the QSL time achieves the actual driving time, i.e., $\tau_{\mathrm{QSL}}=\tau$, there is no potential capacity for further speedup and no speedup can be appear. But, when we have $\tau_{\mathrm{QSL}}<\tau$, it indicates the
potential capacity for quantum dynamical speedup.

We firstly consider the exactly solvable model for a two-level system in the presence of $N-1$ additional atoms (see appendix A).
We suppose that our considered system (the 1th atom) is initially in the excited state and $N-1$ additional atoms are in the ground state and the environment is in a vacuum state.
So, the QSL time bound for the considered atom can be evaluated as
\begin{eqnarray}
\tau_{\mathrm{QSL}}=\frac{\tau(1-|\alpha_{1}(\tau)|^2)}{\int_{0}^{\tau} |\partial_{t}|\alpha_{1}(t)|^{2}|dt},
\end{eqnarray}
where the explicit form of $\alpha_{1}(t)$ is calculated by taking a Lorentzian spectral density for the reservoir in appendix A.
Moreover, in order to study the non-Markovian dynamics of the considered two-level atom, we use a measure of non-Markovianity, which is based on the reverse flow of information from the reservoir back to the system as follows \cite{Breuer}
\begin{eqnarray}
  \Re=max_{\rho_{1,2}(0)} \int_{\vartheta>0} \vartheta[t,\rho_{1,2}(0)]dt,
\end{eqnarray}
where
\begin{eqnarray}
  \vartheta[t,\rho_{1,2}(0)]=\frac{d}{dt} D[\rho_{1}(t),\rho_{2}(t)],
\end{eqnarray}
indicates the changing rate of the trace distance of a pair of states denoted by $D[\rho_{1}(t),\rho_{2}(t)]=\frac{1}{2} Tr|\rho_{1}(t)-\rho_{2}(t)|$ with the trace norm definition for an operator $\chi$ as $|\chi|=\sqrt{\chi^{\dag}\chi}$. For our two-level system, by choosing the optimal pair of initial states obtained in Ref. $\cite{Wissmann}$, it is easy to check that the trace distance of the evolved states can be written as $D[\rho_{1}(t),\rho_{2}(t)]=|\alpha_{1}(t)|^{2}$. Then, one can easily verify that
\begin{eqnarray}
  \Re=\frac{1}{2}\big[ |\alpha_{1}(\tau)|^{2}-1+\int_{0}^{\tau} |\partial_{t}|\alpha_{1}(t)|^{2}|dt \big],
\end{eqnarray}
which connects to Eq. (17) as
\begin{eqnarray}
\tau_{\mathrm{QSL}}=\frac{\tau}{2\frac{\Re}{1-|\alpha_{1}(\tau)|^2}+1}.
\end{eqnarray}
It is clear that, the QSL time is related to the non-Markovianity $\Re$ within the driving time and the atomic excited population $|\alpha_{1}(\tau)|^2$.

In the next step, the QSL time bound for a V-type three-level atom will be obtained in the presence of $N-1$ additional atoms.
To this aim, by considering the exactly solvable model for a V-type three-level atom in the presence of $N-1$ additional atoms (see appendix B)
and assuming the initial condition as $\nu_{1}^{A}(0)=\nu_{1}^{B}(0)=\nu_{1}(0)=2^{-1/2}$ and $\nu_{j}^{A}(0)=\nu_{j}^{B}(0)=\nu_{0}=0$ for $j=2, 3, ..., N$, which implies that $\nu_{1}^{A}(t)=\nu_{1}^{B}(t)=\nu_{1}(t)$, the QSL time can be evaluated as
\begin{eqnarray}
\tau_{\mathrm{QSL}}=\frac{\tau(1-2|\nu_{1}(\tau)|^2)}{2\int_{0}^{\tau} |\partial_{t}|\nu_{1}(t)|^{2}|dt}.
\end{eqnarray}
Also, the measure of non-Markovianty, $\Re$, by considering the optimized pair of initial states for a V-type three-level atom obtained in Ref. $\cite{Gu}$, is reduced to
\begin{eqnarray}
  \Re=2 \int_{\partial_{t}|\nu_{1}(t)|>0} \partial_{t}|\nu_{1}(t)|dt.
\end{eqnarray}
Here, in contrast to the previous case, we emphasize that no explicit relationship can be established between the non-Markovianity and the QSL time bound
of the considered V-type three-level atom.

\section{Results and discussion}
In the following, we illustrate the influence of non-Markovianity and formation of a bound state on the QSL time for our considered model in section III.
In Fig. 2, we present the ratio $\tau_{\mathrm{QSL}}/\tau$ with the behaviors of non-Markovianity of a two-level atom in terms of the coupling strength $\gamma_{0}/\omega_{0}$
and for different numbers of additional atoms in the reservoir. Here, the ratio $\tau_{\mathrm{QSL}}/\tau<1$ indicates the potential capacity for quantum dynamical speedup.
As we can see in Fig. 1, the critical point from no-speedup to speedup of quantum evolution is just the point when the Markovian environment becomes non-Markovian. Also, it is surprisingly seen that although increasing the number of additional atoms leads to more decrement of the QSL time, the non-Markovianity of the system does not always increase by inserting the additional atoms into the reservoir. This means that, non-Markovianity which is induced by the memory effect of environment can not always induce dynamical acceleration in our considered model. To justify it and according to Eq. (21), it should be noted that when $\Re=0$ and provided that $|\alpha_{1}(\tau)|^{2}\neq1$, no-speedup can be observed in the quantum evolution, i.e. $\tau_{\mathrm{QSL}}=\tau$. Moreover, in the absence of additional atoms ($N=1$), and by choosing the actual driving time large enough, no population will be trapped in the excited state of the considered two-level system, i.e. $|\alpha_{1}(\tau)|^{2} \rightarrow 0$. Therefore, the QSL time will be inversely proportional to the non-Markovianity as $\tau_{\mathrm{QSL}}/\tau=({2\Re+1})^{-1}$, and consequently, the non-Markovian effect becomes the unique reason for speeding up of quantum evolution and it induce dynamical acceleration which is matched with results in Ref. $\cite{Deffner}$. However, by inserting the additional atoms into the reservoir ($N>1$), the excited-state population approaches to a non-zero steady value in a large enough time, i.e. $|\alpha_{1}(\tau)|^{2} \rightarrow (N-1)/N$ (see Eq. (A-9) in appendix A). Therefore, the competition between the population, $|\alpha_{1}(\tau)|^{2}$,
and non-Markovianity, $\Re$, takes responsibility for the intrinsic speedup of quantum evolution. On the other hand, we can claim that the reason for the speedup is not solely due to the non-Markovian effect of the reservoir. One question naturally arise: What can be seen as an essential reflection to the quantum speedup?
To answer the question, we further investigate the speedup from the perspective of formation of a system-environment bound state and demonstrate that how the quantum speedup is exclusively related to the formation of the system-environment bound state. The system-environment bound state is actually the stationary state of the entire system where the formation of it can lead to the inhibition of spontaneous emission $\cite{Liu1,Ahansaz1}$. To this aim, the ratio $\tau_{\mathrm{QSL}}/\tau$ with the negative energy spectrum of the total Hamiltonian in Eq. (1), has been sketched in Fig. 3. It is obvious that the transition from no-speedup to speedup of quantum evolution is occurred faster by increasing the number of additional systems.
Also, the critical points for forming the bound state match well with the ones for presenting quantum speedup. It is worth noting that providing stronger bound states in the system-environment spectrum via inserting more additional atoms into the reservoir, can lead to greater speed of quantum evolution.
Thus, we can conclude that the formation of a bound state is the essential reason for the acceleration of evolution.

For more investigation of this interesting finding, we proceed further by considering a V-type three-level atom in the presence of similar and additional atoms.
In this regard, the ratio $\tau_{\mathrm{QSL}}/\tau$ with the non-Markovianity of the V-type three-level atom has been presented in Fig. 4 and with the negative energy spectrum of the total Hamiltonian in Eq. (8), has been presented in Fig. 5. In addition to the previous observations that are also seen for the model considered here, we find that the speedup phenomenon is occurred faster by increasing the SGI parameter. Also, it is clear that increasing the SGI parameter leads to improvement of degree of boundedness, which in turns causes the speedup of the evolution. Consequently, all of these observations returns to the fact that the existence of system-reservoir bound state with higher degree of boundedness ensures the enhancement
in the intrinsic speed of evolution irrespective of non-Markovianity of the system.

\section{Conclusions}
In conclusion, the mechanism of quantum speedup for a two-level system and a V-type three-level system has been studied in the presence of additional systems from two perspectives:
non-Markovianity and formation of a system-environment bound state. We have demonstrated that although non-Markovianity governs the quantum speedup in the absence of additional systems,
but it is neither necessary nor sufficient to speed up the quantum evolution when the environment become complex in the presence of additional systems. Also, it has been revealed that the bound state of the whole system plays a deterministic role in quantum speedup of the considered systems. In the other hand, providing stronger bound states can lead to greater speed of quantum evolution. Our results may open new perspectives for acceleration of evolution through engineering the formation of a bound state.

\newpage

\vspace{1cm} \setcounter{section}{0}
 \setcounter{equation}{0}
 \renewcommand{\theequation}{A-\arabic{equation}}
{\Large{Appendix A:}}\\
\textbf{Dynamics of a two-level system in the presence of $N-1$ additional systems}:
Here, we want to derive the dynamics of an open two-level atomic system in the presence of $N-1$ additional systems. In this regard, we obtain
the exact dynamics of a system consisting of $N$ non-interacting two-level atoms in a common reservoir as considered in section II. 1.
For simplicity, we use the Schrodinger equation in the interaction picture as
\begin{eqnarray}
  i \frac{d}{dt}|\psi(t)\rangle_{I}=H_{I}(t) |\psi(t)\rangle_{I},
\end{eqnarray}
where the Hamiltonian in this picture is given by
\begin{eqnarray}
H_{I}(t)=\sum_{l=1}^{N} \sum_{k} \big(g_{k} {\sigma_{l}^{+}} b_{k} e^{i(\omega_{0}-\omega_{k})t}+g_{k}^{*} {\sigma_{l}^{-}} b_{k}^{\dagger} e^{-i(\omega_{0}-\omega_{k})t}\big),
\end{eqnarray}
and
\begin{eqnarray}
  |\psi(t)\rangle_{I}=\alpha_{0}(t)|0\rangle_{S} |0\rangle_{E}+\sum_{l=1}^{N}\widetilde{\alpha}_{l}(t)|l\rangle_{S} |0\rangle_{E} +\sum_{k}\widetilde{\beta}_{k}(t)|0\rangle_{S} |1_{k}\rangle_{E},
\end{eqnarray}
with $\widetilde{\alpha}_{l}(t)=e^{i\omega_{0}t}\alpha_{l}(t)$ and $\widetilde{\beta}_{k}(t)=e^{i\omega_{k}t}\beta_{k}(t)$.
Substituting Eq. (A-2) and Eq. (A-3) into Eq. (A-1) gives the following differential equations as
\begin{equation}
\begin{array}{l}
\displaystyle \frac{d \widetilde{\alpha}_{l}(t)}{dt}=-i \sum_{k} g_{k} \widetilde{\beta}_{k}(t) e^{i(\omega_{0}-\omega_{k})t},\\\\
\displaystyle \frac{d \widetilde{\beta}_{k}(t)}{dt}=-i \sum_{l=1}^{N} g_{k}^{*} \widetilde{\alpha}_{l}(t) e^{-i(\omega_{0}-\omega_{k})t},
\end{array}
\end{equation}
with $l=1,2,...,N$. It is clear that $H_{I}(t) |0\rangle_{S} \otimes |0\rangle_{E}=0$, so $\alpha_{0}(t)=\alpha_{0}(0)=\alpha_{0}$.
Then, integrating the last equation of Eq. (A-4) and substituting it into the others gives
\begin{eqnarray}
  \frac{d\widetilde{\alpha}_{l}(t)}{dt}=-\int_{0}^{t} f(t-t') \sum_{u=1}^{N}  \widetilde{\alpha}_{u}(t) dt',
\end{eqnarray}
where the correlation function $f(t-t')$ is related to the spectral density $J(\omega)$ of the reservoir by
\begin{eqnarray}
f(t-t')=\int d\omega J(\omega) e^{i(\omega_{0}-\omega)(t-t')}.
\end{eqnarray}
For the reservoir, we take the Lorentzian spectral density as
\begin{eqnarray}
  J(\omega)=\frac{1}{2\pi} \frac{\gamma_{0} \lambda^{2}}{(\omega-\omega_{0})^2+\lambda^{2}},
\end{eqnarray}
where $\omega_{0}$ is the central frequency of the reservoir, the parameter $\lambda$ defines the spectral width of the coupling and $\gamma_{0}$ is the coupling strength.
Therefore, by considering the spectral density $J(\omega)$ given by Eq. (A-7) and using the Laplace transform, the exact solutions of the probability amplitudes $\widetilde{\alpha}_{l}(t)$
can be obtained as $\cite{Ahansaz}$
\begin{eqnarray}
\begin{array}{c}
  \widetilde{\alpha}_{l}(t)=e^{-\lambda t/2}\Big(\mathrm{cosh}{(\frac{Dt}{2})}+\frac{\lambda}{D} \mathrm{sinh}{(\frac{Dt}{2}})\Big) \widetilde{\alpha}_{l}(0)+
  (\frac{(N-1) \widetilde{\alpha}_{l}(0)-\sum_{l'\neq l}^{N} \widetilde{\alpha}_{l'}(0)}{N})\times \\\\
  \Big(1-e^{-\lambda t/2}\big(\mathrm{cosh}{(\frac{Dt}{2})}+\frac{\lambda}{D} \mathrm{sinh}{(\frac{Dt}{2})}\big)\Big),
\end{array}
\end{eqnarray}
where $D=\sqrt{\lambda^{2}-2\gamma_{0} \lambda N}$. Also, by considering the 1th atom as our main concern of system ($l=1$) and assuming the initial condition as $\widetilde{\alpha}_{l'}(0)=0$ for $l'\neq 1$, Eq. (A-8) is reduced to
\begin{eqnarray}
  \widetilde{\alpha}_{1}(t)=\bigg(\frac{N-1}{N}+\frac{e^{-\lambda t/2}}{N}\Big(\mathrm{cosh}{(\frac{Dt}{2})}+\frac{\lambda}{D} \mathrm{sinh}{(\frac{Dt}{2}})\Big)\bigg) \widetilde{\alpha}_{l}(0).
\end{eqnarray}
Moreover, the explicit form of the reduced density operator of the 1th atom in the basis $\{|e\rangle, |g\rangle\}$, can be obtained in the presence of $N-1$ additional atoms by tracing over the reservoir and the atoms except 1th one as follows
\begin{eqnarray}
\rho_{1}(t)=\left(
                \begin{array}{cc}
                  |\widetilde{\alpha}_{1}(t)|^2 &  \alpha_{0}^{*}\widetilde{\alpha}_{1}(t) \\\\
                  \alpha_{0}\widetilde{\alpha}_{1}(t)^{*} & 1-|\widetilde{\alpha}_{1}(t)|^2 \\
                \end{array}
              \right).
\end{eqnarray}

\vspace{1cm} \setcounter{section}{0}
 \setcounter{equation}{0}
 \renewcommand{\theequation}{B-\arabic{equation}}
{\Large{Appendix B:}}\\
\textbf{Dynamics of a V-type three-level system in the presence of $N-1$ additional systems}:
In this appendix, we derive the dynamics of an open V-type three-level atomic system in the presence of $N-1$ additional systems. To this aim, we obtain
the exact dynamics of a system consisting of $N$ non-interacting V-type three-level atoms in a common reservoir as considered in section II. 2.
According to Schr\"{o}dinger equation in the interaction picture, we have
\begin{eqnarray}
i\frac{d}{dt}|\varphi(t)\rangle_{I}=H'_{I}(t)|\varphi(t)\rangle_{I},
\end{eqnarray}
where
\begin{eqnarray}
H'_{I}(t)=\sum_{l=1}^{N} \sum_{m=A,B} \sum_{k} \big(g_{mk} \sigma_{m}^{l+} b_{k}e^{i(\omega_{m}-\omega_{k})t}+g_{mk}^{\ast} \sigma_{m}^{l-} b_{k}^{\dagger}e^{-i(\omega_{m}-\omega_{k})t}\big),
\end{eqnarray}
and
where
\begin{eqnarray}
  |\varphi(t)\rangle_{I}=\nu_{0}(t)|0\rangle_{S} \otimes |0\rangle_{E}+\sum_{l=1}^{N} \big( \widetilde{\nu_{l}^{A}}(t)|A_{l}\rangle+ \widetilde{\nu_{l}^{B}}(t)|B_{l}\rangle \big)_{S} \otimes |0\rangle_{E}
  +\sum_{k} \widetilde{\eta_{k}}(t) |0\rangle_{S} |1_{k}\rangle_{E},
\end{eqnarray}
with $\widetilde{\nu_{l}^{A}}(t)=e^{i\omega_{A}t}\nu_{l}^{A}(t)$, $\widetilde{\nu_{l}^{B}}(t)=e^{i\omega_{B}t}\nu_{l}^{B}(t)$ and
$\widetilde{\eta_{k}}(t)=e^{i\omega_{k}t} \eta_{k}(t)$.
It can be easily seen that $H'_{I}(t) |0\rangle_{S} \otimes |0\rangle_{E}=0$, then $\nu_{0}(t)=\nu_{0}(0)=\nu_{0}$, while by substituting Eqs. (B-2) and (B-3) into Eq. (B-1), we have the following differential equations
$$
\frac{d\widetilde{\nu_{l}^{A}}(t)}{dt}=-i \sum_{k} g_{Ak} e^{i(\omega_{A}-\omega_{k})t} \widetilde{\eta_{k}}(t),
$$
\begin {equation}\label{}
\frac{d\widetilde{\nu_{l}^{B}}(t)}{dt}=-i \sum_{k} g_{Bk} e^{i(\omega_{B}-\omega_{k})t} \widetilde{\eta_{k}}(t),
\end {equation}
$$
\frac{d\widetilde{\eta_{k}}(t)}{dt}=-i \sum_{m=A,B} g_{mk}^{*} e^{-i(\omega_{m}-\omega_{k})t} \sum_{l=1}^{N} \widetilde{\nu_{l}^{m}}(t).
$$
If we integrate the last equation of Eq. (B-4) and substitute it into the others, so the following set of closed integro-differential equations is obtained as
\begin{equation}
\frac{d\widetilde{\nu_{l}^{m}}(t)}{dt}=-\sum_{n=A,B} \int_{0}^{t} f_{mn}(t-t') \sum_{j=1}^{N} \widetilde{\nu_{j}^{n}}(t') dt', \quad m=A, B.
\end{equation}
It should be noted that the kernels in Eq. (B-5) can be expressed in terms of spectral density $J(\omega)$ as follows
\begin{eqnarray}
f_{mn}(t-t')=\int_{0}^{t}d\omega J_{mn}(\omega)e^{i(\omega_{m}-\omega)t-i(\omega_{n}-\omega)t'},
\end{eqnarray}
and $J(\omega)$ is chosen as Lorentzian distribution
\begin{eqnarray}
J_{mn}(\omega)=\frac{1}{2\pi}\frac{\gamma_{mn}\lambda^{2}}{(\omega_{0}-\omega)^{2}+\lambda^{2}},
\end{eqnarray}
where $\omega_{0}$ and $\lambda$ are respectively the center frequency of the structured reservoir and the spectral width of the coupling.
Also, $\gamma_{mm}=\gamma_{m}$ are the relaxation rates of the two upper excited levels and $\gamma_{mn}= \sqrt{\gamma_{m} \gamma_{n}} \theta$ with $m\neq n$ are responsible for the
SGI between the two decay channels $|A\rangle \rightarrow |C\rangle$ and $|B\rangle \rightarrow |C\rangle$ in each atom.
The parameter $\theta$ depends on the relative angle between two dipole moment elements which is related to the mentioned transitions. In the other hand, $\theta=0$ means that the dipole moments of two transitions are perpendicular to each other, which is corresponding to the case that there is no SGI between two decay channels and $\theta=1$ indicating that the two dipole moment are parallel, which is corresponding to the strongest SGI between two decay channels.

If we take Laplace transform from both sides of Eq. (B-5), it becomes
\begin{equation}
p \widetilde{\nu_{l}^{m}}(p)-\widetilde{\nu_{l}^{m}}(0)=-\sum_{n=A,B} \mathcal{L}\{f_{mn}(t)\} \sum_{j=1}^{N} \widetilde{\nu_{j}^{n}}(p), \quad m=A, B.
\end{equation}
Here, $\widetilde{\nu_{l}^{m}}(p)=\mathcal{L} \{\widetilde{\nu_{l}^{m}}(t)\}=\int_{0}^{\infty} \widetilde{\nu_{l}^{m}}(t) e^{-pt} dt$, is the Laplace transform of $\widetilde{\nu_{l}^{m}}(t)$.
Since, the right hand sides of Eq. (B-8) are equal for each $m$, therefore, the following relation between the coefficients is found
\begin{equation}
p \widetilde{\nu_{1}^{m}}(p)-\widetilde{\nu_{1}^{m}}(0)=...=p \widetilde{\nu_{l}^{m}}(p)-\widetilde{\nu_{l}^{m}}(0)=...=p \widetilde{\nu_{N}^{m}}(p)-\widetilde{\nu_{N}^{m}}(0).
\end{equation}
Now, writing the coefficients $\widetilde{\nu_{j}^{m}}(p)$ ($j\neq l$) in terms of $\widetilde{\nu_{l}^{m}}(p)$ and inserting them into the Eq. (B-8), leads to the following equation
\begin{eqnarray}
p \widetilde{\nu_{l}^{m}}(p)-\widetilde{\nu_{l}^{m}}(0)=-\sum_{n=A,B} \mathcal{L}\{f_{mn}(t)\} \bigg(N \widetilde{\nu_{l}^{n}}(p)+\frac{1}{p}\sum_{j\neq l}^{N} \big(\widetilde{\nu_{j}^{n}}(0)-\widetilde{\nu_{l}^{n}}(0)\big) \bigg).
\end{eqnarray}
Here, we consider the case in which the two upper atomic states are degenerated and the atomic transitions are in resonant with the central frequency of the reservoir, i.e
$\omega_{A}=\omega_{B}=\omega_{0}$. Under this consideration, we can assume that $\gamma_{A}=\gamma_{B}=\gamma_{0}$ and $\gamma_{AB}=\gamma_{BA}=\gamma_{0} \theta$, so the kernels in
Eq. (B-6) takes the following simple form
\begin{equation}
\begin{array}{l}
\displaystyle f_{AA}(t)=f_{BB}(t)=f(t)=\int_{0}^{t}d\omega J(\omega)e^{i(\omega_{0}-\omega)(t-t')},\\\\
\displaystyle f_{AB}(t)=f_{BA}(t)=f'(t)=\int_{0}^{t}d\omega J'(\omega)e^{i(\omega_{0}-\omega)(t-t')},
\end{array}
\end{equation}
where $J'(\omega)=\theta J(\omega)$. In the following, we define the new coefficients $\widetilde{\nu_{l}^{\pm}}(p)=\widetilde{\nu_{l}^{A}}(p) \pm \widetilde{\nu_{l}^{B}}(p)$, and rewrite
Eq. (B-10) in terms of them as
\begin{eqnarray}
p \widetilde{\nu_{l}^{\pm}}(p)-\widetilde{\nu_{l}^{\pm}}(0)=-\big[\mathcal{L}\{f(t)\} \pm \mathcal{L}\{f'(t)\} \big] \bigg(N \widetilde{\nu_{l}^{\pm}}(p)+\frac{1}{p}\sum_{j\neq l}^{N} \big(\widetilde{\nu_{j}^{\pm}}(0)-\widetilde{\nu_{l}^{\pm}}(0)\big) \bigg),
\end{eqnarray}
therefore, after taking inverse Laplace transform we get
\begin{eqnarray}
\widetilde{\nu_{l}^{\pm}}(t)=\mathcal{G}_{\pm}(t)\widetilde{\nu_{l}^{\pm}}(0)-\frac{1-\mathcal{G}_{\pm}(t)}{N} \sum_{j\neq l}^{N} \big(\widetilde{\nu_{j}^{\pm}}(0)-\widetilde{\nu_{l}^{\pm}}(0)\big),
\end{eqnarray}
where $\mathcal{G}_{\pm}(t)=e^{-\lambda t/2} \big(\mathrm{cosh}{(\frac{D^{\pm}t}{2})}+\frac{\lambda}{D^{\pm}} \mathrm{sinh}{(\frac{D^{\pm}t}{2})}\big)$ and $D^{\pm}=\sqrt{\lambda^{2}-2\gamma_{0} (1 \pm \theta) \lambda N}$.
Ultimately, the first atom ($l=1$) of the studied system is considered as our main concern of the single three-level system and the $N-1$ remainder ones are considered as the additional atoms. Therefore, the explicit form of the reduced density operator of the $1$th atom in the basis $\{|A_{1}\rangle, |B_{1}\rangle, |C_{1}\rangle\}$, can be obtained in the presence of $N-1$ additional atoms by tracing over the reservoir and the atoms except $1$th one as follows
\begin{eqnarray}
\varrho_{1}(t)=\left(
                \begin{array}{ccc}
                  |\widetilde{\nu_{1}^{A}}(t)|^2 & \widetilde{\nu_{1}^{A}}(t) \widetilde{\nu_{1}^{B}}(t)^{*} & \widetilde{\nu_{1}^{A}}(t) \nu_{0}^{*}\\\\
                  \widetilde{\nu_{1}^{B}}(t) \widetilde{\nu_{1}^{A}}(t)^{*} & |\widetilde{\nu_{1}^{B}}(t)|^2 & \widetilde{\nu_{1}^{B}}(t) \nu_{0}^{*}\\\\
                  \nu_{0} \widetilde{\nu_{1}^{A}}(t)^{*} & \nu_{0} \widetilde{\nu_{1}^{B}}(t)^{*} & 1-|\nu_{1}^{A}(t)|^2-|\nu_{1}^{B}(t)|^2\\
                \end{array}
              \right).
\end{eqnarray}

\newpage
Fig. 1. The figure corresponds to seven atoms ($N=7$) immersed in a common reservoir.
The central atom (the orange circle) is considered as our main concern of the single-atom system and the $N-1$ remainder ones are
considered as the additional atoms (the blue circles).

\begin{figure}
\centering
\includegraphics[width=200 pt]{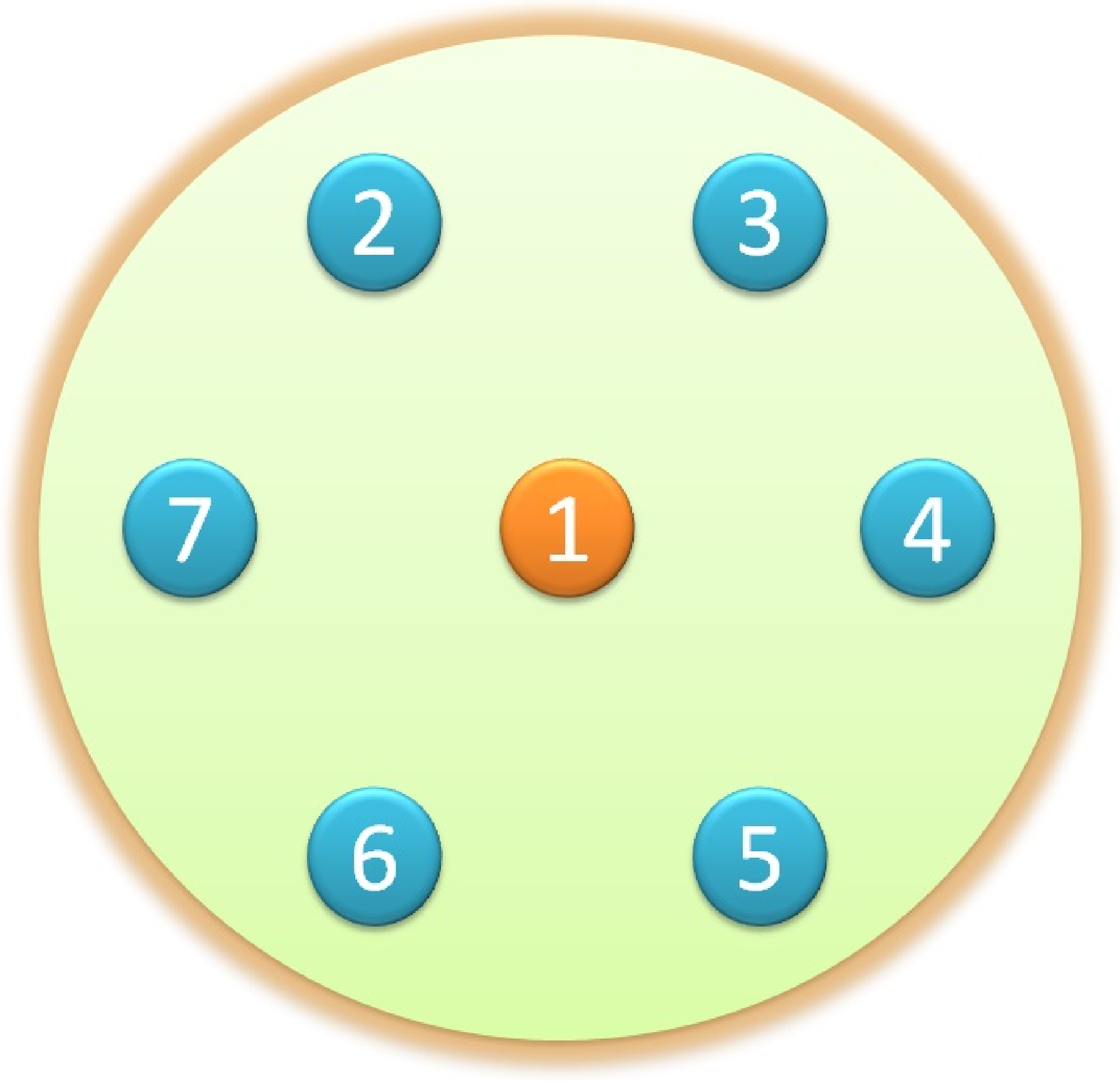}
\caption{}
\end{figure}

\newpage
Fig. 2. The QSL time (blue solid line) and non-Markovianity (red dashed line) for a two level system in terms of the coupling strength $\gamma_{0}/\omega_{0}$ and with different numbers of atoms in the reservoir as (a) $N$=1, (b) $N$=3, (c) $N$=8 and (d) $N$=30. The values of the used parameters are $\lambda=2$ (in units of $\omega_{0}$) and $\tau=5$ (in units of $\omega_{0}^{-1}$).

\begin{figure}
\centering
\includegraphics[width=450 pt]{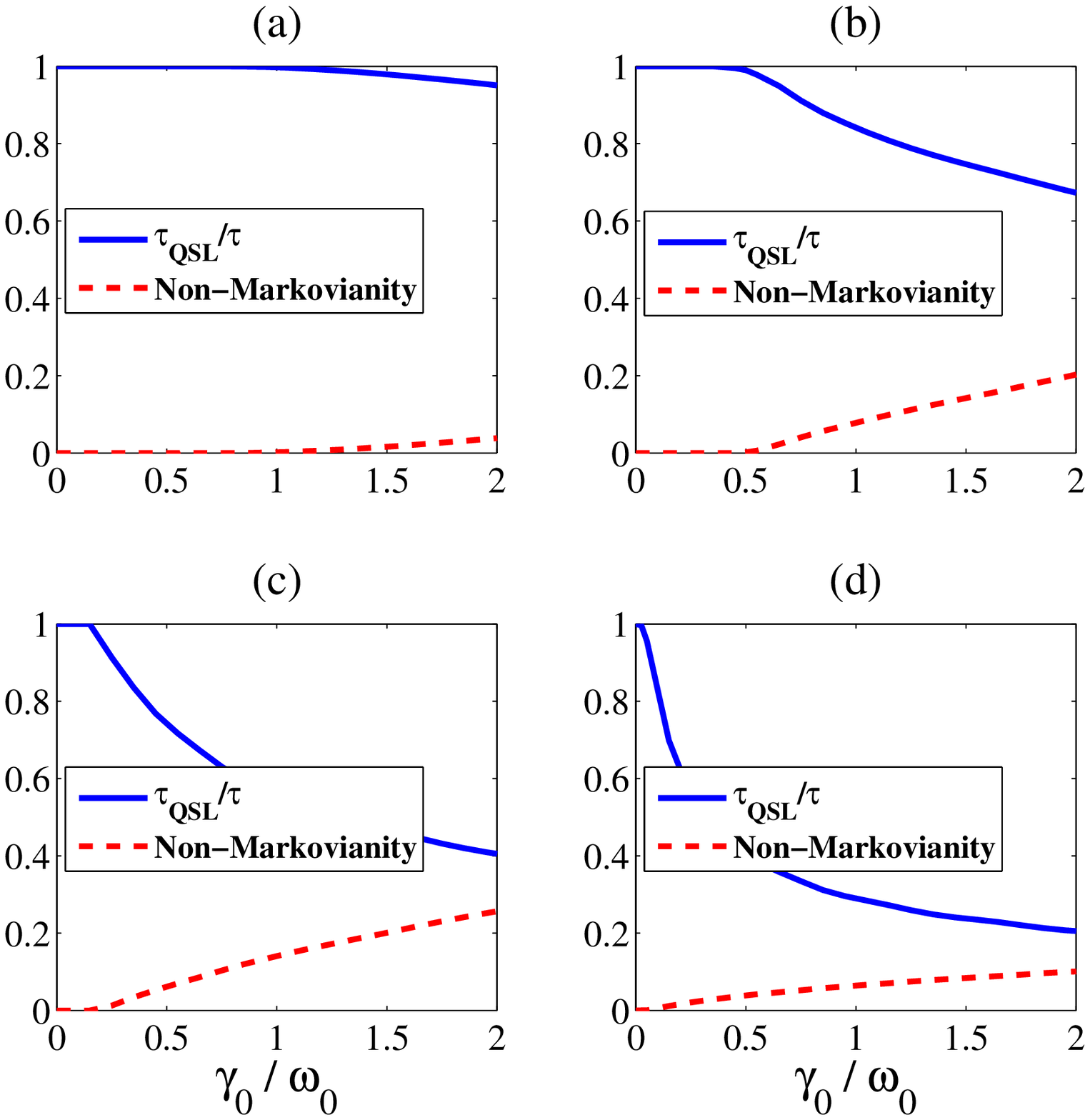}
\caption{}
\end{figure}

\newpage
Fig. 3. The QSL time (blue solid line) for a two level system and the energy of the formed bound state (green dashed line) in terms of the coupling strength $\gamma_{0}/\omega_{0}$ and with different numbers of atoms in the reservoir as (a) $N$=1, (b) $N$=3, (c) $N$=8 and (d) $N$=30. The values of the used parameters are $\lambda=2$ (in units of $\omega_{0}$) and $\tau=5$ (in units of $\omega_{0}^{-1}$).

\begin{figure}
\centering
\includegraphics[width=450 pt]{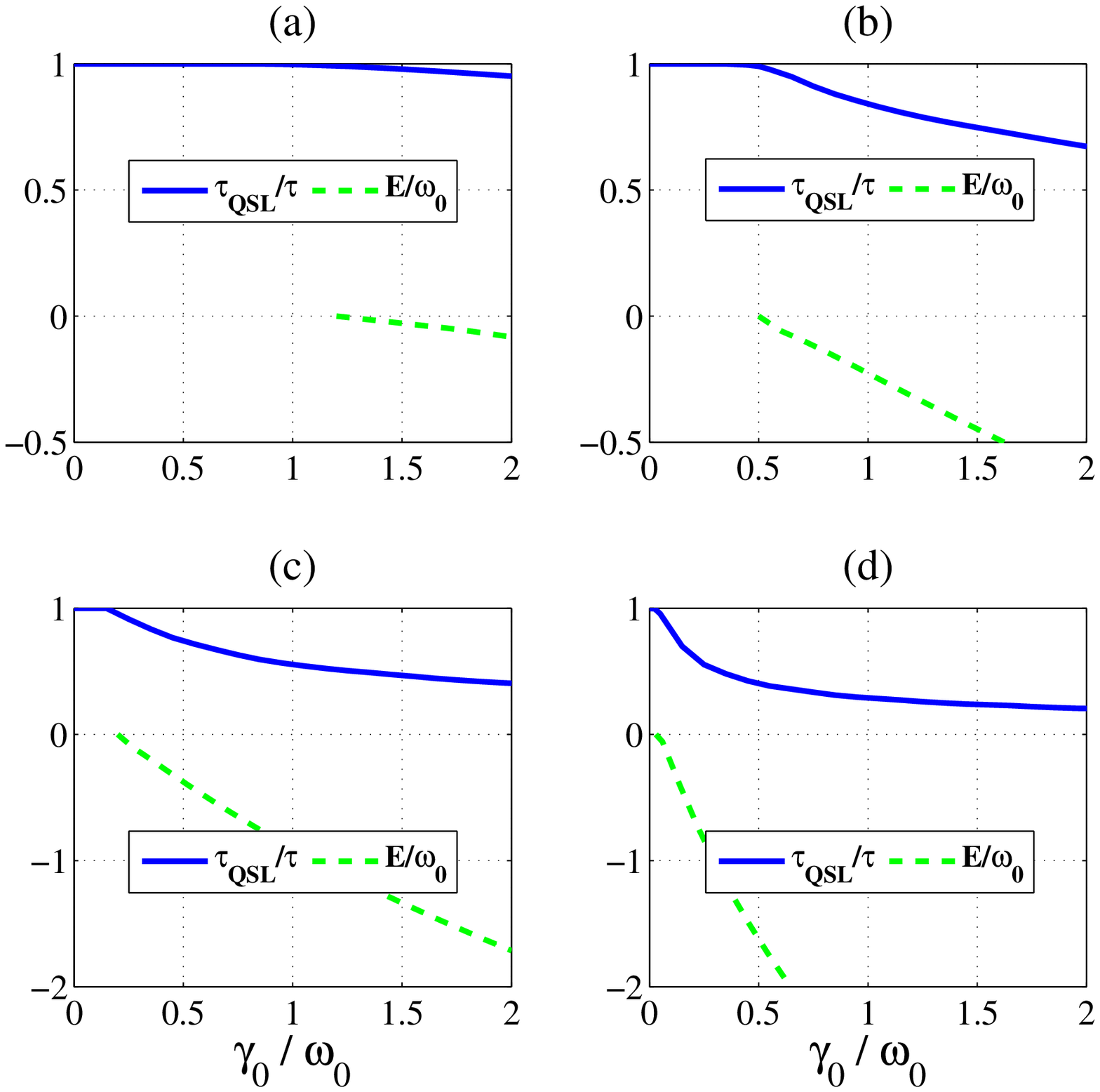}
\caption{}
\end{figure}

\newpage
Fig. 4. The QSL time (blue solid line) and non-Markovianity (red dashed line) for a V-type three level system in terms of the coupling strength $\gamma_{0}/\omega_{0}$ and with different numbers of atoms in the reservoir as (a) $N$=1, (b) $N$=3, (c) $N$=8 and (d) $N$=30. The values of the used parameters are $\lambda=2$ (in units of $\omega_{0}$) and $\tau=5$ (in units of $\omega_{0}^{-1}$). Also, lines with the square marks are plotted for $\theta$=1 and without the marks are plotted for $\theta$=0.

\begin{figure}
\centering
\includegraphics[width=450 pt]{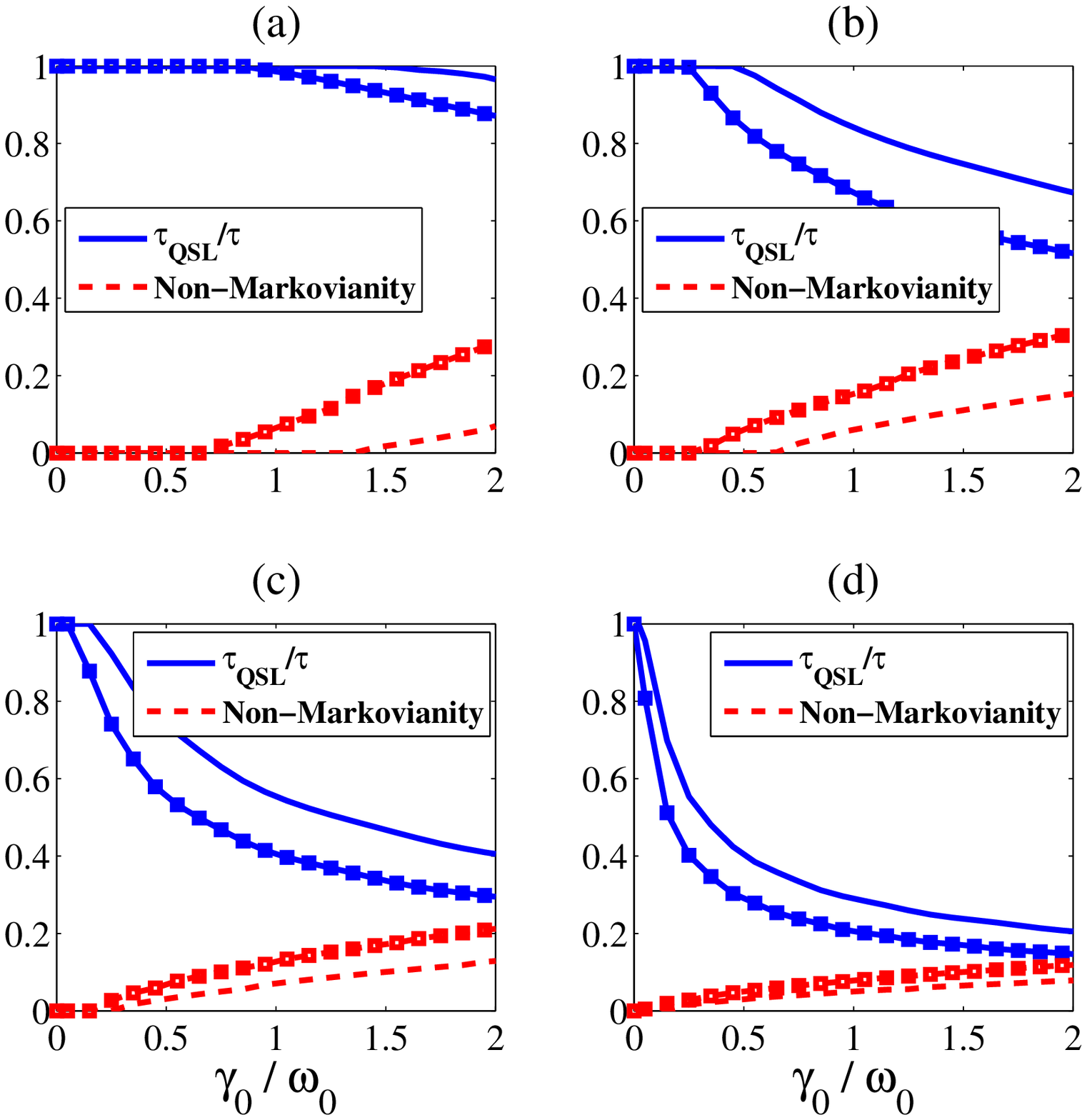}
\caption{}
\end{figure}

\newpage
Fig. 5. The QSL time (blue solid line) for a V-type three level system and the energy of the formed bound state (green dashed line) in terms of the coupling strength $\gamma_{0}/\omega_{0}$ and with different numbers of atoms in the reservoir as (a) $N$=1, (b) $N$=3, (c) $N$=8 and (d) $N$=30. The values of the used parameters are $\lambda=2$ (in units of $\omega_{0}$) and $\tau=5$ (in units of $\omega_{0}^{-1}$). Also, lines with the square marks are plotted for $\theta$=1 and without the marks are plotted for $\theta$=0.

\begin{figure}
\centering
\includegraphics[width=450 pt]{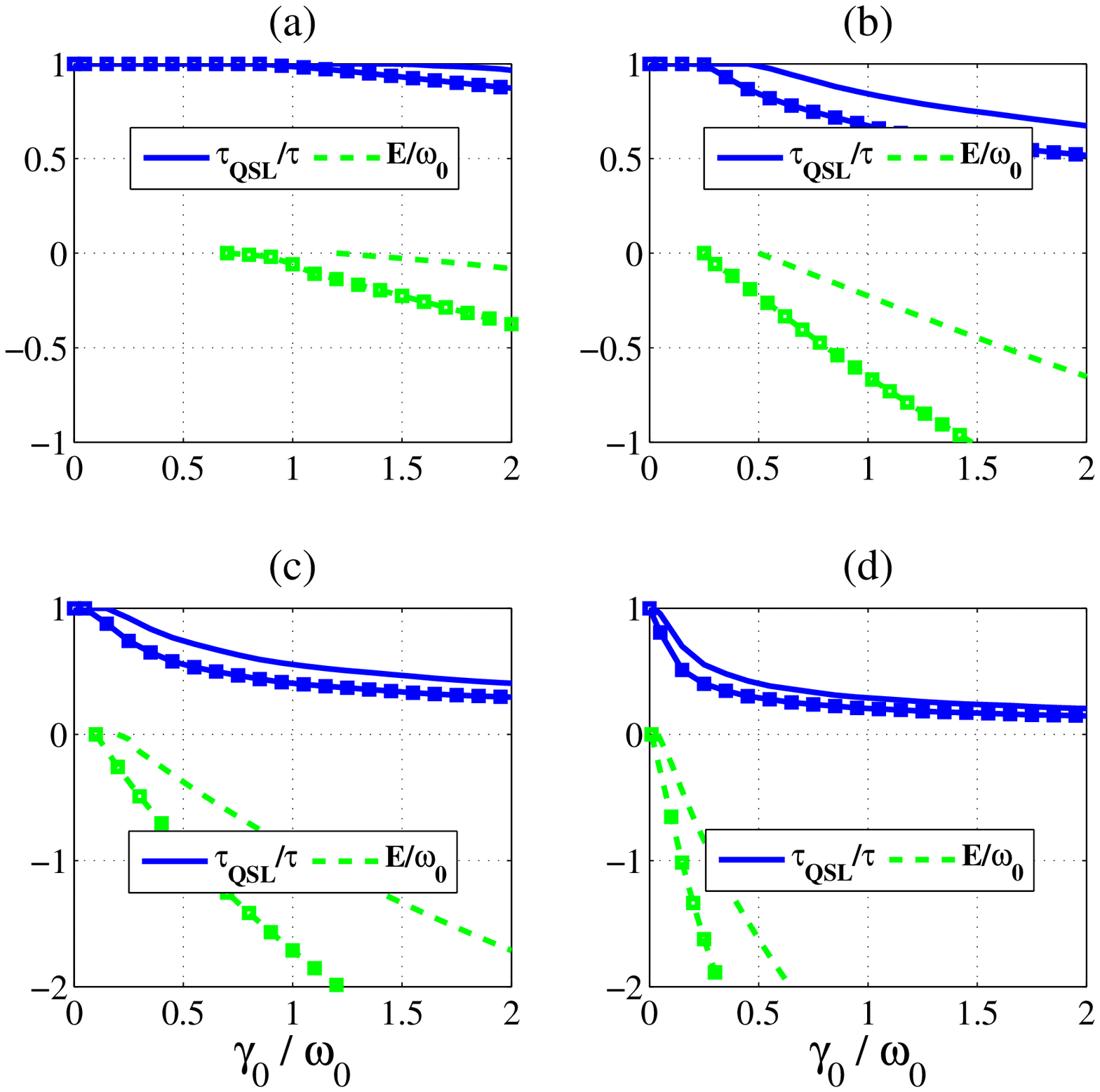}
\caption{}
\end{figure}

\end{document}